\newcommand{\al}{\mbox{$\alpha $}}

\documentstyle[epsf,12pt,epsfig]{article}

\newcommand{\la}{\mbox{$ \lambda $}}
\newcommand{\ls}{\mbox{$ l_{s} $}}

\newcommand{\lj}{\mbox{$ l_{p}^{11} $}}
\newcommand{\be}{\begin{equation}}
\newcommand{\br}{\begin{eqnarray}}
\newcommand{\ee}{\end{equation}}
\newcommand{\er}{\end{eqnarray}}

\newcommand{\p}{\mbox {$ \partial$}}

\begin{document}
\title{
\hfill\parbox{4cm}{\normalsize IMSC/97/12/42 \\hep-th/9712179}\\ 
\vspace{2cm}
 Interaction of F- and D-strings in the Matrix Model} 
\author{N.D. Hari Dass and B. Sathiapalan\\
{\em Institute of Mathematical Sciences}\\
{\em Taramani}\\
{\em Chennai 600113}\\
{\em INDIA }}
\maketitle
\begin{abstract}
We study a configuration of a parallel F- (fundamental) and D-
string in IIB string theory by considering its T-dual configuration 
in the 
matrix model description of M-theory. We show that certain
non-perturbative features of string theory such as 
$O(e^{-\frac{1}{g_{s}}})$
effects due to soliton loops, the existence of bound state (1,1)
strings and manifest S-duality, can be seen in
matrix models. We discuss certain subtleties  that arise in the large-N
limit when membranes are wrapped around compact dimensions.
\end{abstract}
\newpage
\section{Introduction}

 Recently matrix models have been proposed as a non-perturbative 
description of M-theory \cite {BFSS} and of Type IIB string
theory \cite{IKKT,FKKT}. Conceptually the idea is very interesting since 
one can
explicitly write down matrices that represent interesting
configurations such as D-p-branes \cite{BFSS,IKKT,Banks,BSS,Li,Ima,BKS}. 
Quite a bit of work has been
done in extending these ideas as well as in  checking the validity of these models by comparing with
results from perturbative string theory and supergravity
(\cite{sur} - \cite{FKK}).  

	In an earlier paper, one of us \cite{BSSL2} had investigated
certain aspects of the IIB model \cite{IKKT}. A very important property of the 
IIB string theory is its SL(2,Z) duality invariance.  How this
symmetry manifests itself in the IIB matrix model is an open question
although some suggestions were made in \cite{BSSL2}.  Being a 
non-perturbative symmetry, which exchanges F-(fundamental) strings with
D-strings, this is impossible to see directly in the usual perturbative
formulation of string theory.  However the matrix model purports
to be a non-perturbative description and this should be manifest.  In
fact
in the IIA matrix model   \cite{BFSS,BSS}, we can make this manifest 
if we recall
the origin of this symmetry from M-theory \cite{JS1,JS2}.  It corresponds
to interchanging two cycles of a torus on which M-theory can be 
compactified to give IIA theory compactified on $S^{1}$. Since the IIA
and IIB theories are related by T-duality one can then see the SL(2,Z)
symmery of IIB theory.  
      
  Another interesting question is whether one can see the effects
of soliton loops which are expected to give terms of
$O(e^{-1/g_{s}})$\cite{S}. Thus
in the interaction between two D-branes one should expect loops of an
open D-string stretched between the two objets.  In particular, if one
studies an F-string and a D-string, the object that stretches between
them, cannot be an F-string (see Fig. 1). 
\begin{figure}[htb]
\begin{center}
\mbox{\epsfig{file=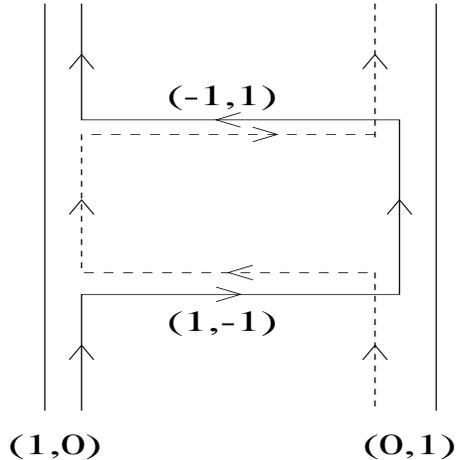,width= 6truecm,height=6truecm,angle=-90}}
\caption{ A time slice of the open string loop diagram. A pair of virtual $(1,-1)$ and
$(-1,1)$ strings connect the external $(1,0)$ and $(0,1)$ strings. Flux
lines are indicated by the dashed and solid lines with arrows on them.}
\label{Fig. 1}
\end{center}
\end{figure}
Flux conservation says that 
it has to 
be a (1,-1) string \cite{JS2},
which can be thought of as a bound state of an F-string  and a
D-string.  Thus, if one can see such effects, on would also be seeing a
bound state \cite{W,DFS,GNS,VS,SS,CP}. 

 In this paper we will try to do the above calculation.  We consider 
an F-string wound around some direction (say, $X^{7}$), and a
D-string,
also wound around the same direction a certain distance away. We
calculate
the interaction between the two (one loop approximation) and look
for the above effects.  This calculation is done in the T-dual IIA
string theory or equivalently in M-theory, using the matrix model of
ref. \cite{BFSS}.
In the M-theory matrix model one can construct such a configuration
in the form of
a membrane (D-2-brane of IIA) wound around $X^{7}$ and $X^{8}$ and
another membrane wound around $X^{9}$ and $X^{7}$.  The latter membrane gives 
an F-string wound around $X^7$,
if $X^{9}$ is taken
to be the ``eleventh '' dimension of M-theory, whose radius fixes the
IIA coupling constant and string tension. Interchanging the eighth
and ninth directions then corresponds to S-duality of the T-dual IIB string
theory. Note that the IIB string theory is the one obtained on
T-dualizing $X^{8}$.  We find that, indeed, non-perturbative effects
described above can be seen. 
    In the process of doing this calculation, we uncover some 
subtleties that arise at one loop, when a membrane is wound around a compact
direction.  This situation has been considered at tree level in \cite{WT,VK}.

This paper is organized as follows. In section II we give a short
review of matrix model formalism and discuss in general terms
a membrane wound around a compact direction.  
 In Section III we focus on
a specific configuration of interest, calculate the one loop
effective action and study the non-perturbative issues mentioned above.
Section IV has some concluding comments.

\section{Generalities}

\subsection{Review of M-theory matrix model}
    The matrix model for M-theory \cite{BFSS} is essentially a
D-0-brane 
action,
reinterpreted as a light cone gauge action where the ``eleventh ''
dimension, of radius $R_{11}$, is part of the light cone coordinates. The
$P^{+}$ component of momentum in the infinite momentum frame is thus 
approximately $\frac{N}{R_{11}}$ (where $N$ is the number of D-0branes),
 and is like a ten dimensional mass. When
$P^{+}$
is large (as it is in an infinite momentum frame) the action is that
of a non relativistic particle with mass $\frac{N}{R_{11}}$.
Thus the evolution operator  $P^{-} = \frac{P_{tr}^{2} +m^{2}}{2P^{+}}$
is the Hamiltonian obtained from the D-0-brane action.
The bosonic part of the action for N D-0-branes\footnote{ This supersymmetric action has also been studied
in a different context by \cite{halp}}is
\be  \label{2.1}
S=
\frac{1}{2g_{s}}\int \frac{dt}{\ls}Tr\{(\p _{t} X^{i})^{2} +
\frac{1}{4\pi ^{2}\ls
^{4}}[X^{i},X^{j}]^{2}\}
\ee

We have explicitly written factors of $\ls$, to make the action
dimensionless.  Our conventions are as follows: $g_{s}$ is the
(IIA) string coupling constant and
we have defined the inverse string tension to be $2\pi \al ' $
with $\al ' = \ls ^{2}$.  The parameters of M-theory are the
radius $R_{11}$ and the eleven dimensional Planck length $\lj $. $\lj$
is defined by the membrane tension which we have taken to be
$\frac{1}{(2\pi )^{2} (\lj ) ^{3}}$.  This fixes $g_{s}^{2} =
(\frac{R_{11}}{\lj })^{3}$ and also $\al ' = \frac{(\lj )^{3}}{R_{11}}$.
These relations also imply that $g_{s} \ls = R_{11}$. Thus the kinetic
term is essentially ``$\frac{1}{2} mv^{2}$'' with $m =\frac{1}{R_{11}}$.
The coefficient of the potential term can be seen to be $\frac{R_{11}}
{8\pi ^{2} (\lj )^{6}}$.
The normalization of the potential term is chosen so that the 
classical mass of
a D-2 brane comes out right. This will become clear as we proceed 
(see equation \ref{2.2.5}).

\subsection{Matrix Formalism}
\setcounter{equation}{0}
A general $N \times N$ matrix can be expanded in a
basis,
\be     \label{3.3}
A = \sum _{m,n} A_{mn}e^{imp}e^{inq}
\ee
where
$q,p$ are $N\times N$ hermitian matrices that satisfy
\be     \label{3.4}
[q,p] = \frac{2\pi i}{N}
\ee                                 
There is the usual caveat that $N$ has to be infinity for this
to work.\footnote{ For a careful treatment of the finite N  analog of this
construction
see \cite{zach}.} To make things concrete, consider a particle in a one
dimensional box of length $L$, with periodic boundary conditions.
Also assume an ultraviolet cutoff i.e. a lattice spacing ``$a$''.
The number of sites is $L/a \equiv N$.
We introduce canonical momentum and position operators $P$ and $Q$.
The momentum eigenfunctions are of the form
\be     \label{3.6}
\phi _{n} \; = \; \frac{1}{\sqrt{L}}e^{ip_{n} x}
\ee
where $p_{n}=\frac{2\pi n}{L} =\frac{2\pi n}{Na}$ is the $n ^{th}$ eigenvalue
of the momentum operator $P$.  Let us choose
$a = \sqrt {\frac{2 \pi}{N}}$ and $L=\sqrt {N2\pi}$.  Then
\be     \label{3.61}
p_{n} = \frac{2\pi n}{\sqrt{N2\pi}}  \: \: \: 0\leq n\leq N-1
\ee
The operator $p=P \sqrt{\frac{2\pi}{N}}$, has eigenvalues 
$\frac{2\pi n}{N}$. The matrix $e^{iP \sqrt{\frac{2\pi}{N}}}$ then is
\be     \label{3.7}
e^{ip} = e^{iP \sqrt{\frac{2\pi}{N}}} =
\left( \begin{array}{ccccc}1 & 0      &   &    &              \\ 
                           0 & \omega &   &    &              \\ 
                             &        &   & 0  &              \\
                             &        & 0 & ...&              \\
                             &        &   &    & \omega ^{N-1}\\  
                            \end{array} \right)
\ee
where $\omega$ are the $N^{th}$ roots of unity.
$Q$ has possible values $0,a,2a,...,na,..(N-1)a$ which is the same
as
\be     \label{3.8}
(0,\frac{1}{\sqrt{N}},.....,\frac{N-1}{\sqrt{N}})\sqrt{2\pi}
\ee
the same as $ P$.  In this normalization $[Q,P]=i$.
If I choose $q=\sqrt{\frac{2 \pi}{N}}Q$ so that
\be     \label{3.10}
[q,p]=\frac{2 \pi i}{N}
\ee 
it has eigenvalues $\frac{n}{N}2\pi$, the same as $ p$.
\be     \label{3.11}
e^{iq}=\left( \begin{array}{ccccc}0 & 1 &  &  &  \\
                            0 & 0 & 1 &  & \\  
                              &   & 0 & 1 & \\
                              &   &   & .. & 1\\
                            1 &   &   &    & 0 \end{array} \right)
\ee
Note again that the matrix $q$ satisfying (\ref{3.10}) and (\ref{3.11})
only exists in the limit $N \rightarrow \infty$ which is also
the limit $L \rightarrow \infty $ or $ a \rightarrow 0$.  In this
limit $q$ and $p$ have continuous eigenvalues ranging from $0$ to
$2\pi$.

  In the one loop calculation we will need to define adjoint action of
  matrices. If we define $\bar {O} = Adjoint (O)$ as
\be	\label{2.1.1}
[O,X] = \bar {O}X
\ee 
Then, clearly
\be
\bar {p} = -\frac{2\pi i}{N} \frac{d}{dq}   \,  \, ;  
\bar {q} = \frac{2\pi i}{N} \frac{d}{dp}
\ee
Note that
\be
[ \bar {p}, \bar {q}] = 0
\ee
Another useful operator is $\sigma _3$ whose adjoint
$\bar {\sigma _3}$ has eigenvalues $\pm 2,0,0$.

The following can also be easily checked:
\be	\label{adj}
 \overline {A \otimes B} = \overline {A}\otimes B + 
A \otimes \overline {B} - \overline {A} \otimes \overline {B}
\ee

\subsection{Membrane wrapped around $S^{1}$}
\setcounter{equation}{0} 

When one space dimension is compactified, a collection of D-0-branes
is physically equivalent to a D-1-brane wrapped around the compact
dimension \cite{JP}.  An explicit construction has been given in
\cite{WT}.  Matrix model compactifications on torii have been
discussed in the literature
\cite{GR,R,LMS,EGKR,VK,BRS,FHRS,AS,NS1}. In this section we
concentrate on the $S^1$ case and investigate what happens at one
loop. This is in preparation for Section 3.

The action is that of  9+1 dimensional Supersymmetric $U(N)$
Yang-Mills theory reduced to 1+1 dimension.
\[	
S=
\frac{1}{2g_{s}}\int \frac{dt}{l_{s}}\int _{0}^{2\pi L_{9}^{*}}
 \frac{dx}{2\pi L_{9}^{*}}
Tr\{ (\p _{t}X^{i})^{2} - (D_{x}X^{i})^{2} + 
\]
\be	\label{2.2.1}
+ (F_{09})^{2}
+\frac{1}{4\pi ^{2}l_{s}^{4}}([X^{i},X^{j}])^{2}\}
\ee

$D_{x} =  \p _{x} +i A_{9}$ is the covariant derivative in a 
direction $X^{9*}$, which is T-dual to $X^{9}$, and is of radius
$L_{9}^{*}$. $x$
is thus a coordinate along a D-1-brane wound around $X^{9*}$.
If $X^{i}_{mn}$ is a matrix describing the lattice of D-0-branes, ($m,n
=0...M$) then
following \cite{WT}, we have
\be	\label{2.2.2}
A^{9} = \frac{1}{2 \pi \al '} \sum _{n=0}^{M} e^{inx\frac{L_{9}}{\alpha '}}
X_{0n}^{9}
\ee 
Note that $X_{00}$ is the original  D-0-brane matrix of the
uncompactified
theory.

Let us now consider a configuration where a membrane of
M-theory, or equivalently a D-2-brane of IIA string theory is wrapped
around this compact dimension.  This can be described as a non-trivial
background for the gauge field in the 1+1 dimensional theory. 
In the D-0-brane language we can describe the membrane
 by using, as membrane coordinates, the $N\times N$ matrices $p,q$  
and the wrapping by \cite{BFSS,BSS} 
\be	\label{2.2.3}
X^{9} = L_{9} p
\ee
The eigenvalues of $p$ range from $0 - 2\pi$, thus (\ref{2.2.3})
describes a membrane with one side wrapped around a circle of radius $L_{9}$.
If we set $X_{00}$ of eqn.(\ref{2.2.2}) to be that given in
(\ref{2.2.3}) we get the background gauge field configuration
corresponding to a wrapped membrane.

We can also wrap the other side of the membrane around another compact
dimension (say $X^{7}$), if we want, by setting
\be	\label{2.2.4}
X^{7} = L_{7} q
\ee
 One can check that the value of the classical Lagrangian given in
 (\ref{2.2.1})
for this membrane
 configuration is 
\be	\label{2.2.5}
\frac{(\frac{L_{7}L_{9}}{\l_{11}^{3}})^{2}}{2\frac{N}{R_{11}}}
\ee
This is equal to 
\be
\frac{(m_{membrane})^2 }{2 p^+}
\ee 
as expected.

Let us consider the spectrum of small fluctuations 
around this background. We will
need to do this in calculating the one loop effective action in a
similar situation in the next section.\footnote{Of course this
particular configuration is a BPS state, hence the one loop action vanishes.}

The relevant operator is
\be
D_x = \p _{x}\otimes I + Adj(I  \otimes  iA^{9})
\ee
Since the $x$-derivative acts on a different space than the matrix
indices
we have used a direct product notation for clarity. Also 
since $A^{9}$ acts by commutation we have used the adjoint notation
introduced in the last sub-section. The space of eigenfunctions that
it is acting on is of the type
\be
e^{ir\frac{x}{L_{9}^{*}}}e^{imp}e^{inq}
\ee
Here $r,n,m$ are integers.
Using the fact that $adj(I\otimes p) = I\otimes \frac{-2\pi i}{N}\p _{q}$
we find that the eigenvalues are $ \frac{rN + n}{NL_{9}^{*}}$.
\begin{figure}[htb]
\begin{center}
\mbox{\epsfig{file=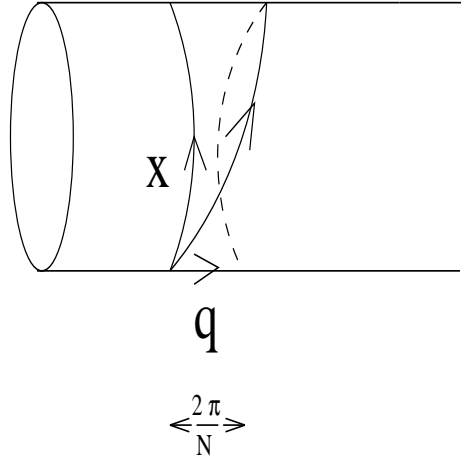,width= 6truecm,height=6truecm,angle=-90}}
\caption{
The translation operator 
$\p _x \otimes I + I \otimes \frac{1}{N}\p _q$ moves along the spiral shown here.}
\label{Fig. 2}
\end{center}
\end{figure}
 
Note that the covariant derivative $D_x$ has become
\be
\p _x \otimes I + I \otimes \frac{1}{N}\p _q
\ee
While $x$ is a coordinate along the compact direction, in the usual
interpretation where a wrapped membrane gives a string, $q$
is a coordinate along the string (which is also along the unwrapped
direction of the membrane).  Thus the operator $D_x$ generates a translation
along a spiral, as shown in Fig 2. Thus a rotation by $2\pi$ along
the $x$ direction is accompanied by a shift $\frac{2\pi}{N}$ in
the transverse direction.  Only when it winds $N$ times around does
it come back to the starting point.

 Thus the effective (dual) radius in the presence of a wrapped membrane is
$NL_{9}^{*}$, rather than $L_{9}^{*}$!  This is reminiscent of the 
``long string '' phenomenon that has cropped up in different places
\cite{DM,MS,Motl,Ver}.  We will see this same effect in the next
section where we discuss an F-string and a D-string.

\section{F-D interaction}
\setcounter{equation}{0}
\subsection{The Configuration}

We consider M-theory with three compact directions with radii $R_{11},
L_{9},L_{8}$. $R_{11}$ is the radius of what we have thus far referred
to as the eleventh dimension and defines a IIA string theory with
 an inverse
tension $\al ' = \frac{(\lj )^{3}}{R_{11}}$.  But one can  equally
well 
consider a IIA 
string theory defined by an inverse tension
$\beta ' = \frac{(\lj )^{3}}{L_{9}}$. Let us refer to the first
description as the $\al '$  description and the second one as
the $\beta '$ description.  Our strategy will be to
 define a matrix model
using the $\al '$ description 
and perform calculations involving two membranes wrapped
around $L_{9}$ and $L_{8}$ respectively.  
In this description
these are 
two wrapped
D-2-branes.  On the other hand, in the $\beta '$ description
one of these (the one wrapped around $X^9$)
becomes the fundamental or F-string and the other is a wrapped
D-2-brane.  We can now consider the T-dual of the $\beta
'$-description with
the duality being performed on $X^{8}$ (i.e. the dual radius
$\tilde L_{8} = \frac{\beta '}{L_{8}}$).  Now our configuration is 
 a D-1-brane (or a D-string) and an F-string of IIB theory. 
Thus in the T-dual $\beta'$ desription we have the
configuration that we wanted
 viz, an  F-string and a
D-string.  For convenience we can wind these two around a large
compact coordinate $X^{7}$ of radius $L_{7}$. We will assume $L_{7}$
is large enough that the winding sectors are unimportant for the
dynamics. We will place these two strings at a distance $b$ from each
other in the $X^{6}$ direction.  All the calculations will however
be performed using the matrix model describing D-0-branes of the $\al '$
description. Thus, to summarize, in the $\al '$ description we have two
D-2-branes wrapped around $L_{9},L_{7}$ and $L_{8},L_{7}$
respectively. We calculate the one loop effective potential using the
matrix model. In the end we will interpret the result in terms of the 
T-dual of the $\beta '$ description.

The string coupling constant in the $\al '$ description is 
$g_{s\alpha '}^{2}=(\frac{R_{11}}{\lj })^{3}$ while in the 
$\beta '$ description
it is $g_{s\beta '}^{2}=(\frac{L_{9}}{\lj })^{3}$.  When we T-dualise, it
becomes $\tilde g_{s\beta '} = g_{s\beta '}\frac{\tilde L_{8}}{\ls } = 
\frac{L_{9}}{L_{8}}$.
This is the usual relation between M-theory on $T^{2}$ and 
IIB string theory \cite{JS1,JS2,HW}. 

As explained in the introduction, this is a configuration where
one should be able to see S-duality.  S-duality merely interchanges
the F-and D-strings and inverts the coupling constant. Furthermore 
the open string connecting an F-string to a D-string cannot be an
F-string. In fact the simplest possibility is a (1,-1) string which
is expected to exist and can also be considered to be a bound state of a
(1,0) and a (0,-1) string.  Thus in effect we have soliton loops and
one expects terms of $O(e^{-\frac{1}{g_s}})$ in the effective action.
These are the two aspects that we would like to see explicitly. So let
us proceed with the calculation.

\subsection{One loop effective action}

Following section 2 we can describe the above configuration by
the following $2N \times 2N$ matrices:
\begin{eqnarray}	\label{3.1.1}
X^{9}_{0,0} &=& \left( \begin{array}{cc} L_{9}p & 0 \\
				0 & 0 \end{array}\right)\nonumber \\
X^{8}_{0,0}&=&\left( \begin{array}{cc} 0 & 0 \\
				0 & L_{8}p \end{array}\right)\nonumber
				\\
X^{7}_{0,0}&=&\left( \begin{array}{cc}L_{7}q  & 0 \\
				0 & L_{7}q \end{array}\right)\nonumber
				\\
X^{6}_{0,0}&=&\left( \begin{array}{cc} \frac{b}{2}  & 0 \\
				0 & -\frac{b}{2} \end{array}\right)
\end{eqnarray}

The subscripts on the matrices represent the additional indices
introduced by the compactification of the 8th and 9th directions
as described in \cite{WT}
The action is now a 2+1 SYM theory on radii $L_{8}^{*},L_{9}^{*}$. 
Although $X^{7}$ is compact, since we have assumed that $L_{7}$ is
large
enough that the winding sector does not contribute to the dynamics, we
do not write a 3+1 SYM theory. The
string length is defined by $\al '$ as before. (Remember that all
calculations
are in the $\al '$ description.)
It is ($i,j$ :1-7)
\[
\frac{1}{g_{s}}\int \frac{dt}{l_{s}}
\int _{0}^{2\pi L_{9}^{*}} \frac{dy}{2\pi L_{9}^{*}}
\int _{0}^{2\pi L_{8}^{*}} \frac{dx}{2\pi L_{8}^{*}}
Tr\{ ( \p _{t}X^{i})^{2} - (D_{x}X^{i})^{2} - (D_{y}X^{i})^{2} + 
\]
\be	\label{3.1.2}
+ (F_{09})^{2}+ (F_{08})^{2}+ (F_{89})^{2}
+\frac{1}{4 \pi ^2 l_{s}^{4}}([X^{i},X^{j}])^{2}\}
\ee
     
Note that $g_{s}L_{8}^{*}L_{9}^{*}=g_{s}^{*}\ls ^{2}$, where $g_s^*$
 is the coupling
constant of the IIA theory after two T-dualities in the
8th and 9th directions \footnote{The reader should not confuse this
with $\tilde g_{s}$ which was defined as the IIB coupling obtained after
a T-duality along $X^8$ in the $\beta '$ description}. 
The covariant derivatives are as in section 2:
$D_{x}= \p _{x} +i A_{8}$ and $D_{y}= \p _{y} +i A_{9}$. 
Equation (\ref{3.1.1}) implies that $A_{8,9}$ have 
expectation values which are
\[
<A_{8}>=\frac{1}{2\pi \al '} X^{8}_{0,0}
\]
\be
<A_{9}>=\frac{1}{2\pi \al '} X^{9}_{0,0}
\ee
In addition $X^{6},X^{7}$ have expectation values.

In calculating
the one loop effective action we have to identify a part of the action
that is quadratic in small fluctuations about a background. 
This can be done as in \cite{IKKT}. The result is that
the fluctuation operator for the bosonic fluctuations is
\begin{equation}
{\cal H}_{bos} = P_{\lambda}^2\delta_{\mu\nu} -2iF_{\mu\nu}
\end{equation}
Here $P_{\la}$ are (adjoint) operators,
and are
defined below.
We will assume that they all have dimensions of momentum. And,
\be
F_{\mu \nu} = i[P_{\mu},P_{ \nu}]
\ee
is also an adjoint operator.

Let
\begin{eqnarray}
p_{6} &=& \frac{1}{2\pi \al '}X^{6}_{0,0} \nonumber \\
 p_{7} &=& \frac{1}{2\pi \al '}X^{7}_{0,0} \nonumber \\
p_{8}&=& <A_{8}> \nonumber \\
p_{9}&=& <A_{9}> 
\end{eqnarray}
Then,
\begin{eqnarray}	\label{3.1.4}
P_{6}&=& I \otimes Adj (p_{6}) \nonumber \\
P_{7} &=& I \otimes  Adj (p_{7})\nonumber \\
P_{8} &=&  -i\p_{x} \otimes I   + I \otimes Adj (p_{8})\nonumber \\
P_{9} &=&  -i\p_{y} \otimes I + I \otimes Adj (p_{9})\nonumber \\
P_{0} &=&  -i\p_{t}\otimes I
\end{eqnarray}
We have split the space into direct product of functions of $x,y,t$ on
which the derivatives act 
and the space of matrices on which the adjoint of the matrices act.

We will now make a rotation of the axes in the (8,9) plane into
$(8',9')$ such that $f_{79'}$ is proportional to the identity. This will
make $F_{79'}=0$. The required transformation is
\begin{eqnarray}
p_9^{\prime} & = & {L_9p_8+L_8p_9\over \sqrt{L_8^2+L_9^2}}\nonumber\\
p_8^{\prime} & = & {L_{9}p_{9}-L_{8}p_{8}\over \sqrt{L_8^2+L_9^2}}
\end{eqnarray}
This gives (we have set $\al ' =1$) using (\ref{3.1.1}) and in an
obvious notation where $2N\times 2N$ matrices are expressed as a direct
product of an $N\times N$ matrix and a $2\times 2$ matrix,
\begin{eqnarray}
p_9^{\prime} & = &{1\over 2\pi\sqrt{L_8^2+L_9^2}}
[L_8L_9  p \otimes I ]\nonumber\\
p_8^{\prime} & = &{1\over 2\pi \sqrt{L_8^2+L_9^2}}[
{L_9^2-L_8^2\over 2}    p \otimes I + 
{L_8^2+L_9^2\over 2}    p\otimes \sigma_3 ~]
\end{eqnarray}
Using (\ref{adj}) we can calculate the adjoints of these and substitute
  into (\ref{3.1.4}). We get (in units where $\al ' =1$)
\begin{eqnarray}
P^{'}_{8}&=& {-i \over \sqrt{L_8^2+L_9^2}} (L_9\p _y - L_8 \p _x )\otimes
I \otimes I + I\otimes
 P_8^{'I} + I \otimes P_8^{'II} \nonumber \\
P_8^{'I} &=& {\sqrt {L_8^2 +L_9^2} \over 4\pi }
i(-\frac{2\pi i}{N}\frac{\p}{\p
q}\otimes \sigma _3 + p \otimes \bar {\sigma _3} + \frac{2 \pi
i}{N}\frac{\p}{\p q} \otimes \bar {\sigma _3} )\nonumber \\
{P^{\prime}}_8^{II}&=& 
i{L_9^2-L_8^2\over 4\pi \sqrt{L_8^2+L_9^2}}(-\frac{2\pi i}{N} 
\frac{\p}{\p q}\otimes I)
\end{eqnarray}
\be
P_{9}^{'} = {-i \over\sqrt {L_8^2 +L_9^2}}(L_9 \p _x + L_8 \p _y)\otimes
I + {L_8 L_9 \over 
\sqrt{L_8^2 +L_9^2}}(\frac{ -i}{N} \frac{\p}{\p q} \otimes
I)
\ee
\be
P_7 = I\otimes L_7(\frac{ i }{N} \frac {\p}{\p p} \otimes I)
\ee
\be
P_6= I\otimes \frac{b}{4\pi}I \otimes \overline {\sigma _3}
\ee
One can also check the following commutators:
\[ 
[P_8 ' , P_9 ']=[P_7 , P_9 ']= 0
\]
\be
i[P_7 ,P_8 ' ] = F_{78^{\prime}}=-{\pi L_7\sqrt{L_8^2+L_9^2}\over 4\pi
^2 N}\overline {\sigma_3}
\ee
$P_6$ commutes with all of them.  The non-zero eigenvalues of 
$F_{78'}$ are thus $\pm \omega = \pm { L_7\sqrt{L_8^2+L_9^2}\over
2 \pi N}$.  
Thus on
this subspace $P_7$ and $P_{8'}$ behave like harmonic oscillator
variables and one concludes that the eigenvalues of
$P_7^2+P_{8^{\prime}}^2$ 
are given by
\begin{equation}
E_{\bar n} = {4\pi L_7\over N}\sqrt{L_8^2+L_9^2}(\bar n+{1\over 2})
=2\omega (\bar n + \frac{1}{2})
\end{equation}
 These operators act on a space spanned by the matrix functions:
\footnote{These are to be multiplied by a $2\times 2$ matrix on which
$\bar {\sigma _3}$ gives 2.}
\begin{equation}
\Phi_{mnrs} = e^{imp}~e^{inq}~e^{ir{x\over L_8^*}}~e^{is{y\over L_9^*}}
\end{equation}
The harmonic oscillator eigenfunctions are to be made by taking linear
combinations of $e^{imp}$.  If we denote these by $H_{\bar n}(p)$, then
the eigenfunctions 
$H_{\bar n}(p)e^{inq}e^{ir{x\over L_8^*}}~e^{is{y\over L_9^*}}$ for
any value of $r,s,n$ have the same eigenvalue $E_{\bar n}$.

$P_{9'}$ has eigenfunctions $\Phi _{mnrs}$ for any values of
$m,n,r,s$.  Its eigenvalues are 
\[
\tilde p_9 = { L_8L_9\over\sqrt{L_8^2+L_9^2}}[(r+s)
+{ n \over N}]
\]
If we define 
\be
R = {\sqrt{L_8^2+L_9^2}\over L_8L_9}=\sqrt{L_8^{*2} +L_{9}^{*2}}
\ee
we see that the eigenvalues are $\frac{1}{NR}[N(r+s) + n]$. The range
of $n$ is 1 - $N$ and so we see that the combination $N(r+s) +n$ runs
over the entire range $1$ to $ NM$ without repetition. (We have assumed
that the combination $ (r+s)$ and $r-s$ have ranges from 1 to M
wher $M$ is some cutoff). Thus the only
combination of integers that does not appear in the expression for
the eigenvalues is $(r-s)$, thus leading to a
degeneracy of $O(M)$
\footnote{Terms involving $(r-s)$ have the effect of translating
``$p$'' 
so that
$H_{\bar n}(p)$ becomes $H_{\bar n}(p- const)$. This does not 
affect $E_{\bar n}$.}.  To summarize 
\begin{equation}
{\cal E} = P_0^2+(\frac{b}{2\pi \alpha '})^2+(2\bar n +1)\omega +
(\frac{N(r+s)+n}{NR})^2
\end{equation}
are the eigenvalues of $P_{\la }^{2}$ 
(in the subspace where $F_{78'}$ is non-zero) 
and the degeneracy  of each eigenvalue is M. Note that $NR$ is thus
the 
effective radius of the dual theory and has acquired an extra factor N
just as we saw in section 2.  Finally the $P_0$ eigenfunctions are of
the form $e^{ip_0 t}$. We can assume a spacing of $\frac{1}{T}$ for
the
eigenvalues  $p_0$, where $T$ is an infrared cutoff in the time
direction.  
This brings a factor of $T$ when converting the trace over eigenvalues
of $P_0$ to an integral over $p_0$.  Finally, the fermionic
determinant
also involves the same operators \cite{IKKT}.

The calculation of the one loop effective action is straightforward
once we have all the above information.
 The final answer for the one-loop effective action is
\be	\label{3.3.1}
T2M \int \frac{dt}{t} \int dp_{0} \sum _{v}
exp\{- \frac{(p_{0}^{2}+ (\frac{v}{NR})^{2} + (\frac{b}{2\pi \alpha
'})^{2})}
{\omega}t \}
2tanh (\frac{t}{4}) sinh ^{2} (\frac{t}{4})
\ee
The coefficient of T can be interpreted as the potential energy
of the configuration.  The various factors of $M,N$ in (\ref{3.3.1})
imply that they are parameters
of the theory that  have some physical significance. 
If we are to take the limit $M,N \rightarrow \infty$ one has 
to worry about finiteness of the final answer.  Furthermore some terms
that describe physical effects may go to zero.  Let us therefore
first investigate the physics of the formula (\ref{3.3.1}) for
finite $N,M$.  Later we will describe
a renormalization of the other parameters that eliminates all $M,N$
dependence. 

We can consider the limit where $b$ is much larger than all other
lengths $L_{7},L_{8},L_{9}, \ls$. Thus we can consider $ t$
to be small and expand the hyperbolic function, which starts off as
$t^{3}$. We also use a Poisson resummation to extract systematically 
the effects of a finite $R$.  We find the following:
\be
V = \frac{M \omega ^{3} 3 \pi}{64}[\frac{4NR}{3b^{4}} + \sum
_{m=1}^{\infty}
\frac{8 m^{2}\pi ^{2} (NR)^{3}}{3b^{2}}K_{-2}(2bm\pi NR)]
\ee
\be	\label{POT}
=\frac{M \omega ^{3} 3 \pi}{64}[\frac{4NR(\al ')^{4}}{3b^{4}} + \sum
_{m=1}^{\infty}
\frac{8 m^{2}\pi ^{2} (NR)^{3}(\al ')^{2}}{3b^{2}}\sqrt {\frac{
\al'}{4mbNR}}
e^{-\frac{mb\pi NR}{\alpha '}}\{1+ O(1/b)\}]
\ee
where we have inserted appropriate powers of $\al '$.

The $b^{-4}$ is as expected when there are three compact dimensions.
The power of $b^{-\frac{5}{2}}$ for the Yukawa terms is also
as expected and corresponds to a simple pole in momentum space.
As explained earlier, this configuration can be viewed, either as two
D-2-branes wound around $X^{8}, X^{7}$ and $X^{9},X^{7}$ or in the
T-dual picture as an
F-string and a D-string, both wound around $X^{7}$.  
In the latter case the string coupling $\tilde g_{s\beta '}$ 
of the IIB theory 
is defined as  
$\frac{L_{9}}{L_{8}}$, and the string tension by $\frac{1}{\beta '}=
\frac{L_{9}}{(\lj)^{3}}$.
Thus $\frac{NR}{\alpha '}$ which can be written as $N\sqrt{L_{8}^{*2} +
L_{9}^{*2}}/\al '$ can now be rexpressed as
$\frac{N\tilde{L_{8}}}{\beta '}\sqrt{1+\frac{1}{\tilde {g_{s\beta
'}}^{2}}}$, 
where 
$\tilde{L_{8}}$ is the dual of
$L_{8}$ with respect to $\beta '$ (i.e. $\tilde L_8 = \frac{\beta '}{L_8}$).  
This is the mass of a (1,1)(or (1,-1)) string wound around 
a radius of $N\tilde{L_{8}}$!.
Thus the Yukawa terms in the potential 
can be understood as the
contribution (in the closed string channel)
of closed (1,-1) strings of the IIB theory wound a radius
$N\tilde{L_{8}}$. The fact that there is a simple pole (in momentum
space) proves the existence of a bound state.  These terms are also
the $O(e^{-\frac{1}{g_{s}}})$ terms that one expects in
non-perturbative string theory.

The
value of $\tilde {L_{8}}$ in terms of the original M-theory parameters
is $\frac {(\lj )^{3}}{L_{9}L_{8}}$. This is the usual connection
between
M-theory on a 2-torus and IIB on a circle \cite{JS1,HW}. The factor $N$
comes from the presence of the wrapped membrane.  It is reminiscent
of a ``long string'' . Somehow a wrapped membrane has additional
low energy excitations that make it look as if the radius of
compactification (in this case the dual radius) is N times larger.  In
the limit $N\rightarrow \infty$ these terms drop out.  However
as will be shown later, it is possible to renormalize the parameters
such that this $N$ dependence disappears.

We thus see that the Yukawa terms represent the 
non-perturbative $e^{-\frac{1}{g_{s}}}$
effects that are expected in string theory due to soliton loops. Here
the soliton is the (1,-1) open string connecting a D-string with an
F-string. The fact that a (1,-1) string  connects an F-string
with a D-string is also expected when you consider flux conservation 
\cite{JS2}.  
 In fact, as mentioned above, one can view the above calculation as a 
demonstration of the existence of a  (1,-1) string.  

Another point worth mentioning is that S-duality (which consists
of interchanging $L_{8}$ and $L_{9}$) is manifest in this formalism.

Finally, there are a couple of simple checks to see that the answer
(\ref{POT}) is correct. The long distance force between a
 D-string and an anti-D-string
should be exactly twice that between an F-string and a D-string (when
they are degenerate in mass, i.e. $\tilde g_{s\beta '}=1$).  
This is because the
$B^{ij}$ contributes to the force in one case but not in the other.
 If one repeats the calculation with two
membranes
wound around $X^{9}$ after setting $L_{9}=L_{8}$ one finds that
the net effect is to replace $\omega$ by $\sqrt {2} \omega$ and
$R$ by $\frac{R}{\sqrt{2}}$.  Making these changes in the leading part
of \ref{POT} we see that we get a factor of 2.

The other check is to compare the answer with what one expects for the
gravitational potential between two objects in Newtonian gravity.
We consider the case where $L_{8}=L_{9}$.  The mass of a 
D-2-brane is $m=\frac{L_{8}L_{7}}{g_{s}l_{s}^{3}}$.  The gravitational
coupling constant is $G \approx \frac{\l_{s}^{8}g_{s}^{2}}{L_{7}L_{8}L_{9}}$.
The potential is thus 
\be
G\frac{m^{2}}{b^{4}} \approx
\frac{l_{s}^{2}L_{7}}{b^{4}}
\ee. 

We can also do this in the T-dual $\beta '$ description, pretending that we
are in a IIB theory and, of course, get the same answer. Note that the
answer
is manifestly T-dual symmetric when we perform T-duality in  $X^{8}$ and
$X^{9}$.  The configuration we started off with was manifestly symmetric
under this T-duality, hence the final expression should also be so.
To compare this with (\ref{POT}) we have to map this potential to
an M-theory calculation as was done for instance in \cite{LM}.  In
M-theory matrix model we calculate $\approx
\frac{m_{int}^{2}}{P_{int}^{11}}$, where $m_{int}$ is the contribution
of the interaction to the rest mass and $P^{11}_{int}$ is the momentum
due to this mass. This is $\approx \frac{m_{int}}{\gamma }$, where
$\gamma$ is the boost factor.  As shown in \cite{LM} there is an extra
factor of $\gamma$ due to the fact that in the boosted frame the
distance between clusters of D-0-brane shrinks, and so we get a
larger answer. So we get two powers of $\gamma$:
$\frac{m_{int}}{\gamma^{2}}$.  
We can use
\be 
\gamma = \frac{P^{11}}{\frac{L_{8}L_{7}}{g_{s}\ls ^{3}}}
\ee
\be
= \frac{N (\lj )^3 }{R_{11}L_8 L_7} = \frac{N \ls ^2}{L_8 L_7}
\ee

 Here we have used $P^{11} = \frac{N}{R_{11}}$. Putting all this 
together we get
\be
\frac{L_{7}^{3}L_{9}^{2}R_{11}}{N^{2}b^{4}\lj ^{3}}
\ee 
One can check that \ref{POT} gives this answer (except for the factor
of $M$ which can be absorbed in $T$).  Note that 
the limit $N \rightarrow \infty$ does not affect the result for
the potential energy. It just makes the boost factor $\gamma$ infinite.
However it does affect the Yukawa terms.

One can also renormalize the parameters $L_7 , L_8 , 
L_9 , b, R_{11} , \lj $ such that physical quantities 
such as
a) rest mass of the membrane 
 , and b) the potential energy calculated in the rest frame, 
have no $N,M$
dependences.  The rescaling is done by requiring that all the 
dependence on $N,M$ are in the boost factors $\gamma $ for the
non interacting piece (\ref{2.2.5}) and $\gamma ^2$ for the
interacting piece (\ref{POT}). Thus the physics becomes $N,M$ independent.
In fact we find that even the renormalized $\gamma$ can be made
$N,M$ independent, so nothing depends on $N,M$. This is intriguing.
Perhaps when additional processes are considered the situation will
be clarified.  The rescaling is as follows:(Barred quantities
are held fixed as $N,M \rightarrow \infty$)
\[	
L_{8,9}=NM \bar L_{8,9}
\]
\[
L_{7}=M \bar {L_{7}}
\]
\[
R_{11} = N \bar R_{11}
\]
\[
b= M\bar {b}
\]
\be
(\lj )^3 = NM^2 (\bar l_p^{11} )^3
\ee
The above rescalings also imply that
\[
\al ' = M^2 \bar {\al ' }
\]
One can check that if we plug in these rescalings, the final answer
has no $N,M$ dependence.  The masses of the membranes (which were
finite and $N,M$ independent to begin with) are unchanged by this
rescaling. The exponent in the Yukawa potential can be seen to be
finite: $\frac{\bar R\bar b}{\bar {\alpha '}}$.  Other processes
need to be investigated before we fully understand the significance of this
rescaling.

\section{Conclusions}

In this paper we have calculated the one loop effective action for 
a configuration of an F-string and a parallel D-string some distance
away. The motivation was to see, within the context of the matrix
model, whether non-perturbative effects mentioned in the introduction 
viz, S-duality and $O(e^{-\frac{1}{g_{s}}})$ effects can be seen.
The calculation demonstrates that it is indeed possible. In fact the
soliton that produces this effect are the (1,-1) strings and thus this
calculation also is an independent proof of the existence of these
bound states.

The intriguing factor of $N$ in (\ref{3.3.1})  needs an explanation.  
Somehow wrapping a membrane has introduced light excitations that
make the dual radius much larger. This does not affect the physics
as far as the leading term is concerned.  But in higher order terms
the exponential fall-off with distance is affected by $N$.
Presumably analogous phenomena
exist on higher dimensional torii.

The significance of the renormalization is also a little unclear.
If we are given that freedom, the significance of the parameter
$N$ seems to disappear. One has to check whether this renormalization
can be consistently done for all processes.

Finally it would be interesting to repeat this calculation in the
IKKT \cite{IKKT} matrix model for the IIB superstring.

\noindent
{\bf Acknowledgements}\\
We would like to acknowledge many discussions with T. Jayaraman who
collaborated with us in the early stages.  Part of this work was done
while the authors were visiting the Centre for Theoretical Studies
at the Indian Institute of Science, Bangalore. We would like to
thank Professor J. Pasupathy and members of CTS for their hospitality.

\end{document}